\documentclass[12pt]{article}
\usepackage{cite}
\usepackage[margin = 1in]{geometry}
\usepackage[colorlinks=true]{hyperref}
\usepackage{graphicx} 
\usepackage[compat=1.1.0]{tikz-feynman}
\usepackage{tikz}
\usepackage{indentfirst}
\usepackage{placeins}
\usepackage{slashed}
\usepackage{gensymb}
\usepackage{amsmath}
\usepackage{amssymb}
\usepackage{multirow}
\usepackage[utf8]{inputenc}
\usepackage{adjustbox}

\newcommand\pubnumber{}
\newcommand\pubdate{\today}

\def\napoli{
    Institute for Mathematics, Astrophysics and Particle Physics\\
    Radboud University, 6525 AJ Nijmegen, Netherlands\\
    The Bethe Center for Theoretical Physics\\
    Bonn University, 53115 Bonn, Germany\\
}

\def\Title#1{\begin{center} {\Large #1 } \end{center}}
\def\Author#1{\begin{center}{ \sc #1} \end{center}}
\def\Address#1{\begin{center}{ \it #1} \end{center}}

\newcommand\pubblock{\rightline{\begin{tabular}{l} \pubnumber\\
         \pubdate  \end{tabular}}}
\newenvironment{Abstract}{\begin{quotation}  }{\end{quotation}}

\begin{document}

\def\thefootnote{\fnsymbol{footnote}}
\def\gsim{\:\raisebox{-0.5ex}{$\stackrel{\textstyle>}{\sim}$}\:}
\def\lsim{\:\raisebox{-0.5ex}{$\stackrel{\textstyle<}{\sim}$}\:}
\def\mET{\slashed{E_T}}

\begin{titlepage}
\pubblock

\vfill \Title{Multi--lepton Signatures from $U(1)_{L_\mu-L_\tau}$ at the LHC}
\Author{Jochem Kip\footnote{jochem.kip@ru.nl},
        Zhongyi Zhang\footnote{zhongyi@th.physik.uni-bonn.de}}
\Address{\napoli}
\vfill
\begin{Abstract} 
The $U(1)_{L_\mu-L_\tau}$ extended Standard Model (SM) is anomaly free, and contains a massive $Z^\prime$ boson. The associated Higgs, which generates the $Z'$'s mass via spontaneous symmetry breaking (SSB) can mix with the SM Higgs. The new parameters relating to extra Higgs cannot be probed at the LHC with final states containing no more than $4$ leptons. Therefore, we use signatures with at least $6$ leptons to probe the parameter space through LHC experiments. Since SM predicts a lower cross section for a final state with at least $6$ leptons than is currently visible at the LHC, the background is negligible, thereby making this channel an extremely sensitive probe. We find that in a limited region of the parameter space this channel can strongly constrain the associated $U(1)_{L_\mu-L_\tau}$ coupling constant, even more so than for a $4$ lepton or fewer final state.
\end{Abstract}
\vfill

\end{titlepage}
\setcounter{footnote}{0}
\section{Introduction}

Currently one of the biggest questions in particle physics is the exact nature of Dark Matter (DM). One model that allows for the incorporation of DM is an extension of the Standard Model with a local $U(1)_{L_\mu-L_\tau}$ gauge symmetry~\cite{Lmu-Ltau_first_paper:PhysRevD.44.2118}, where a DM candidate can take the form of a complex scalar~\cite{Scalar_DM:Baek_2016} or Dirac fermion~\cite{DREES2019130}. While the inclusion of an additional $U(1)$ gauge symmetry in the SM will in general induce anomalies that must be dealt with by choosing the fermionic charges appropriately~\cite{Anomaly_1:PhysRev.177.2426, Anomaly_2:PhysRev.184.1848}, the $U(1)_{L_\mu-L_\tau}$ extension is automatically anomaly free~\cite{Anomaly_cancel_1:PhysRevD.43.R22, Anomaly_cancel_2:Ma_2002}. Additionally, since the newly introduced gauge boson does not couple (directly) to electrons and positrons, strong constraints from $e^+e^-$ colliders such as LEP are avoided which therefore still allows for a much larger parameter space and correspondingly a lower mass spectrum for the newly introduced particles.

One of the best experiments to currently probe the potential existence of such a model and its properties is the LHC. Previous analyses have already investigated large regions of the parameter space that are in reach of the LHC and previous colliders for the $U(1)_{L_\mu-L_\tau}$ model~\cite{Lmu-Ltau_coll_1:Baek_2009, Lmu-Ltau_coll_2:Bandyopadhyay_2009, Lmu-Ltau_LEP1:Carena_2004, Lmu-Ltau_LEP2:Cacciapaglia_2006, Lmu-Ltau_LHC-1:Aad_2014, DREES2019130,U(1)_Bounds:A:2023wup, LHC_search_U1:Heeck:2011wj, LCH_search:Komachenko:1989qn}. Here we propose and investigate a new channel, namely that of a 6-8 lepton final state, whose cross section is too small in the SM to be currently observable, thereby providing a very low SM-produced background. This channel thus allows for the detection of comparatively small cross sections due to the cleanness of the signal.

This paper is structured as follows. First we discuss the relevant theoretical aspects of the model and the details of the 6-8 lepton final state in Section~\ref{sec:2}. Subsequently, we show our results for the cross section and validate that this region of the parameter space is not already excluded by Higgs decay in Section~\ref{sec:3}. We finish with our conclusions in Section~\ref{sec:4}.

\section{Lagrangian and Signatures}
\label{sec:2}
By extending the SM with a local $U(1)_{L_\mu-L_\tau}$ gauge symmetry a new gauge boson $Z'$ is introduced. The resulting field strength tensor and covariant derivative are defined in the usual way
\begin{align}
    Z^\prime_{\mu\nu} = \partial_\mu Z^\prime_\nu - \partial_\nu Z^\prime_\mu && D_\mu = \partial_\mu - i g_{\mu\tau}qZ^\prime_\mu\,.
\end{align}
where $g_{\mu\tau}$ is the associated coupling constant of $U(1)_{L_\mu - L_\tau}$ and $q$ the $L_\mu - L_\tau$ charge of the particle on which the coviarant derivative acts.

Naturally this new gauge boson needs to eat a Goldstone boson to acquire mass. In order to do so a new complex Higgs singlet with non-zero $L_\mu-L_\tau$ charge is introduced which provides mass to the $Z^\prime$ boson. Here we set the $L_\mu - L_\tau$ charge of the new Higgs singlet to be 1. Setting $\phi_h$ to be a SM-like Higgs doublet and $\phi_H$ the new complex Higgs singlet, the additional Higgs terms as compared to the SM are then
\begin{align}
    &\mathcal{L}_H = (D_\mu \phi_H)^*(D^\mu \phi_H)- V(\phi_h, \phi_H)\,,\\
    &V(\phi_h, \phi_H) = \mu^2_H\phi_H^*\phi_H +	\lambda_H(\phi_H^*\phi_H)^2 + \lambda_{hH}(\phi_h^\dagger\phi_h)(\phi_H^*\phi_H)\,.
\end{align}
Here $\mu_H^2$ and $\lambda_H$ are the mass and interaction term of $\phi_H$ with $\mu_H^2 < 0$. After spontaneous symmetry breaking the Higgs doublet and singlet will have the following form in the unitary gauge
\begin{align}
\phi_h = \begin{pmatrix} 0 \\ \frac{v+h}{\sqrt{2}}\end{pmatrix} && \phi_H = \frac{v_{\mu\tau} + H}{\sqrt{2}}\,.
\end{align}
where $v_{\mu\tau}$ is the vacuum expectation value of $\phi_H$. Due to the presence of $\lambda_{hH}$ a mixing is introduced between $h$ and $H$ resulting in two mass eigenstates $h_1$ and $h_2$, of which one must be the 125 GeV SM-like Higgs. We define $h_1$ to be the SM-like Higgs. In order to parameterize the mixing of the Higgs states $h$ and $H$ into $h_1$ and $h_2$ we use a mixing angle $\alpha$ ($\tan(2\alpha) = vv_{\mu\tau}\lambda_{hH}/(\lambda_hv^2-\lambda_Hv_{\mu\tau}^2)$) which defined such that $h_1=h$ and $h_2 = H$ for $\alpha = 0$. The phenomenological implications of $\alpha$ is more straightforward when compared to $\lambda_{hH}$; the size of $\alpha$ dictates the strength of the coupling of massive SM particles to $h_2$, which would result in noticeable effects in SM observations for a large $\alpha$. Therefore, $\alpha$ always needs to be kept sufficiently small, a constraint that we implement in this study by only considering $\alpha = 0.01$ and $0.1$.

Two relevant new vertices that are introduced by the $U(1)_{L_\mu-L_\tau}$ extension are:
\begin{align}
\adjustbox{valign=c}{
\includegraphics[width = 0.2\textwidth]{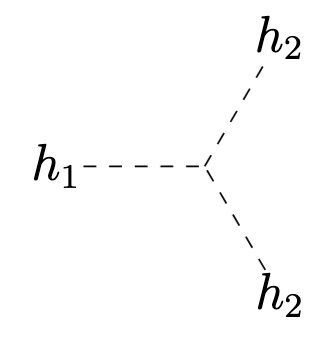}
}
    = -\sin(2\alpha) g_{\mu\tau}\frac{M_{h_1}^2 - M_{h_2}^2}{4M_{Z^\prime}}\,,
    &&
\adjustbox{valign=c}{
\includegraphics[width=0.2\textwidth]{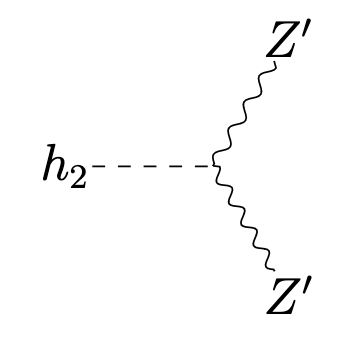}
}
    = g_{\mu\tau}M_{Z^\prime}\,.
    \label{eq:feynman_rules}
\end{align}
After considering these 2 new vertices, all the particles in the model could acquire masses through spontaneous symmetry breaking from Higgs mechanism, instead of merely a simplified model. The simplified models only consider renormalizable terms with testable sectors, and neglects the potential source of the terms. In analyses focusing on final states with up to $4l$, these 2 vertices are not testable.

This model can also accommodate a Dark Matter particle as a complex singlet and provide a mechanism for neutrino masses in a type-I seesaw manner. However, in this study we will focus on probing the mixing angle $\alpha$ of physical Higgs states at the LHC, we shall thus omit from discussing both the Dark Matter and neutrino sectors further.

In order to probe $\alpha$ at the LHC we consider a 6 or 8 lepton final state. Currently this process is invisible at the LHC for the SM (The cross sections of more than 6 lepton final states are smaller than $10^{-5}$ pb. This is lower than the minimal value according to the recent detection limit, i.e. observing at least 3 events under $139$fb$^{-1}$). This is therefore a clean low-background channel, making detection very efficient. The dominant diagram that is introduced by the $U(1)_{L_\mu-L_\tau}$ extension of the SM is

\begin{center}
\includegraphics[width = .6\textwidth]{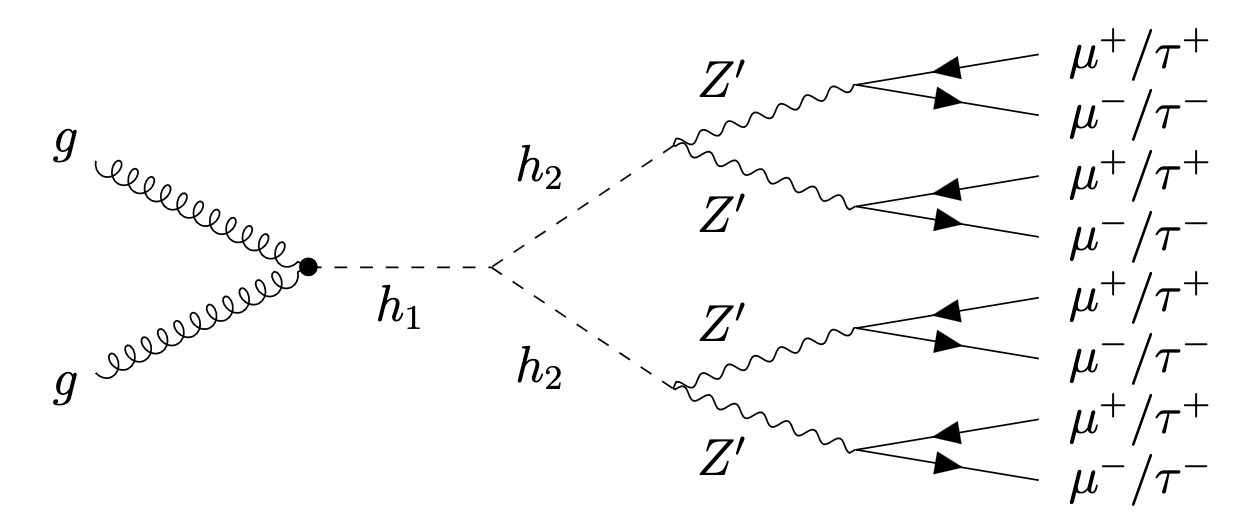}
\end{center}
where the $ggh_1$ vertex is of course an effective vertex induced by quark loops, most notably the top quark~\cite{ggh_crosssec:https://doi.org/10.23731/cyrm-2017-002}. Furthermore, here any $\mu^\pm \tau^\pm$ pair can become a $\nu_\mu/\nu_\tau$ pair thereby making this process a 0,2,4,6, or 8 observable lepton final state. We of course focus only on the 6/8 $\mu/\tau$ final state. From equation \eqref{eq:feynman_rules} it can be seen that the $h_1h_2h_2$ coupling depends on $\alpha$, which is what makes this channel sensitive to the Higgs mixing angle. It should of course also be noted that the cross section from this diagram scales roughly as $g_{\mu\tau}^6$ due to the 3 vertices involving either $h_2$ or $Z^\prime$. Naturally there are many more diagrams contributing to a 6/8 lepton final state, but all of these are subleading with respect to the aforementioned diagram. The reason is that this diagram contains 3 BSM vertices, which are collectively proportional to $g_{\mu\tau}^3$,\footnote{The $g_{\mu\tau}^4$ factor of the 4 $Z'$-lepton-lepton vertices does not contribute here, since the $Z'$ can only decay into leptons.} and 1 $ggh$ vertex. In contrast, other leading order diagrams have at least 3 BSM vertices and 2 electroweak vertices. Additionally, contributions arising from leptons radiating off a $Z'$ are suppressed with a factor of $1/M_{Z'}^2$ as compared to the $h_2$-mediated case per $Z'$ pair. Furthermore, electroweak bosons do not contribute significantly, as the $Z'$ boson does not mix with any of the SM vector bosons.

In order to simulate events we use {\tt MadGraph}~\cite{MadGraph1:Alwall_2014, MadGraph2:Alwall_2015} in which the model files for the $U(1)_{L_\mu-L_\tau}$ are custom made and can be obtained from the authors. Notably, these model files include an explicit $ggh_1$ coupling in order to include the loop-induced gluon fusion to Higgs at tree level. The inclusion of this vertex is vital as all computation performed here are at tree level and Higgs production is foundational to probing both $\alpha$ and the dominant 6/8 lepton final state channel. Furthermore, the number of diagrams scales approximately as $n!$ for $n$ external particles~\cite{kleiss2021}, thus we limit the number of simulated diagrams in order to address the otherwise exceedingly high computation time. We simulate $pp \to Z^\prime Z^\prime Z^\prime Z^\prime$ in which either every $Z^\prime$ goes to a $\mu^+ \mu^-/\tau^+ \tau^-$ pair, or one $Z^\prime$ goes to a $\nu_\mu \overline{\nu}_\mu/\nu_\tau\overline{\nu}_\tau$ pair with all others going to $\mu^+\mu^-/\tau^+\tau^-$, thereby assuring either an 8 or 6 charged lepton final state.

\section{Bounds from LHC Experiments}
\label{sec:3}
In order to set the value of the effective gluon-gluon-Higgs vertex the process $gg \to h$ is simulated in {\tt MadGraph} and the coupling is tuned until the cross section is 48.5$\pm$ 0.1 pb \cite{ggh_crosssec:https://doi.org/10.23731/cyrm-2017-002}. This results in $g_{ggh} = 0.356$.

\begin{figure}[h]
    \centering
    \includegraphics[width = 0.8\textwidth]{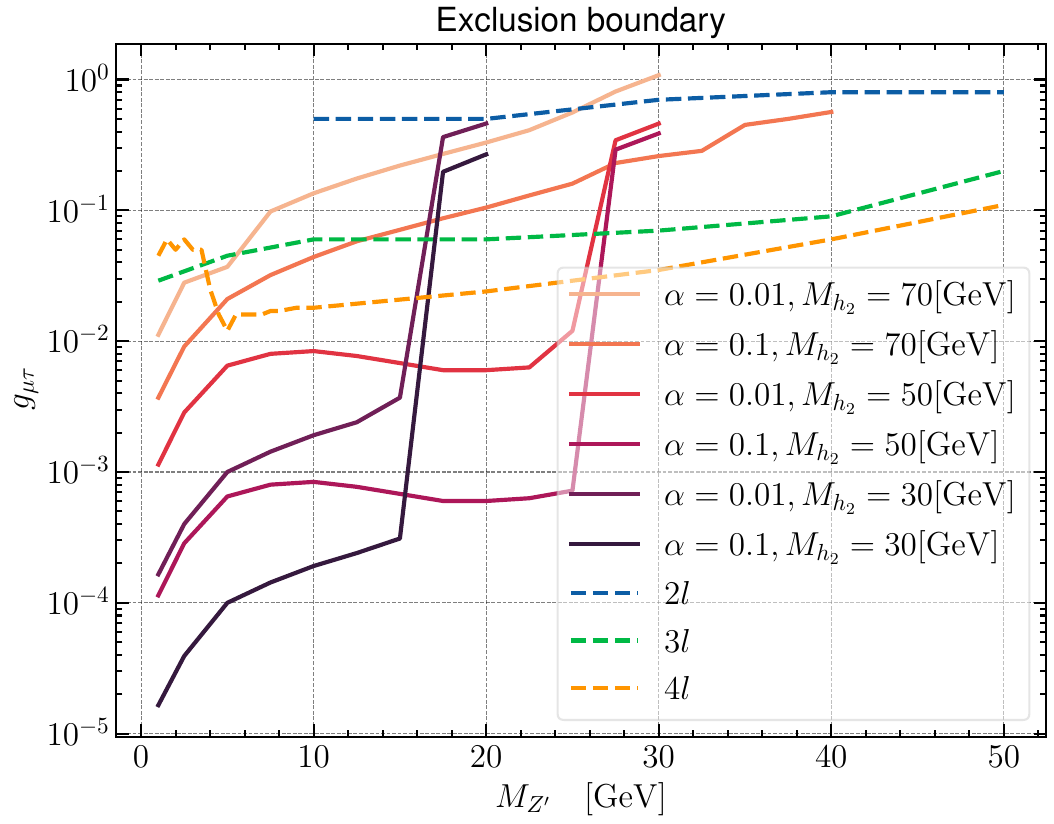}
    \caption{The value of the coupling constant $g_{\mu\tau}$ for which the cross section of the 6/8 lepton process is in the range $2.2 \cdot 10^{-5} \pm 0.11 \cdot 10^{-5}$ pb is on the vertical axis and the mass of the $Z'$ on the horizontal axis. The exclusion boundary of the 6/8 lepton search for all six combinations of $\alpha = 0.1$ and 0.01, and $M_{h_2} = 30$, 50, and 70 GeV is given, in addition to the exclusion boundaries of the 2, 3, and 4 lepton searches as a comparison.}
    \label{fig:exclusion_lines}
\end{figure}

In order to find the values of $g_{\mu\tau}$ that provide exclusion based on the 6/8 lepton channel we consider cross sections larger than $2.2\cdot 10^{-5}$ pb to be excluded. Figure~\ref{fig:exclusion_lines} shows the exclusion lines for all combinations of $\alpha = 0.1$, $0.01$ and $M_{h_2} = 30$, $50$, and $70$ GeV. The values of the exclusion boundary of the 2/3/4 lepton search from \cite{234_lep_search:https://doi.org/10.48550/arxiv.2109.07674} is provided as a comparison. 

From figure~\ref{fig:exclusion_lines} it can clearly be seen that the 6/8 lepton search outperforms the 2/3/4 leptop search in the regions where $M_{Z'} \leq 1/2 M_{h_2}$ and $M_{h_2} \leq 1/2 M_{h}$. This is in line with expectations, seeing as when the on-shell decay $h_2\to Z' Z'$ is not possible if $2 M_{Z'} > M_{h_2}$, and similarly $h_1 \to h_2 h_2$ cannot occur on shell if $2 M_{h_2} > M_{h_1}$. As expected the exclusion boundary for $\alpha =0.1$ is higher than those of $\alpha =0.01$ if all other parameters are equal, differing roughly by one order of magnitude if $M_{h_2} < 1/2 M_{h_1}$. Naturally, since for $M_{h_2} > 1/2 M_{h_1}$ $h_1\to h_2 h_2$ can never be on-shell the sharp rise of the exclusion boundary for $M_{h_2} = 30$ GeV at $M_{Z'} = 15$ GeV and $M_{h_2} = 50$ GeV at $M_{Z'} = 25$ GeV is absent for $M_{h_2} = 70$ GeV. 

\begin{figure}[h]
    \centering
    \includegraphics[width = 0.3\textwidth]{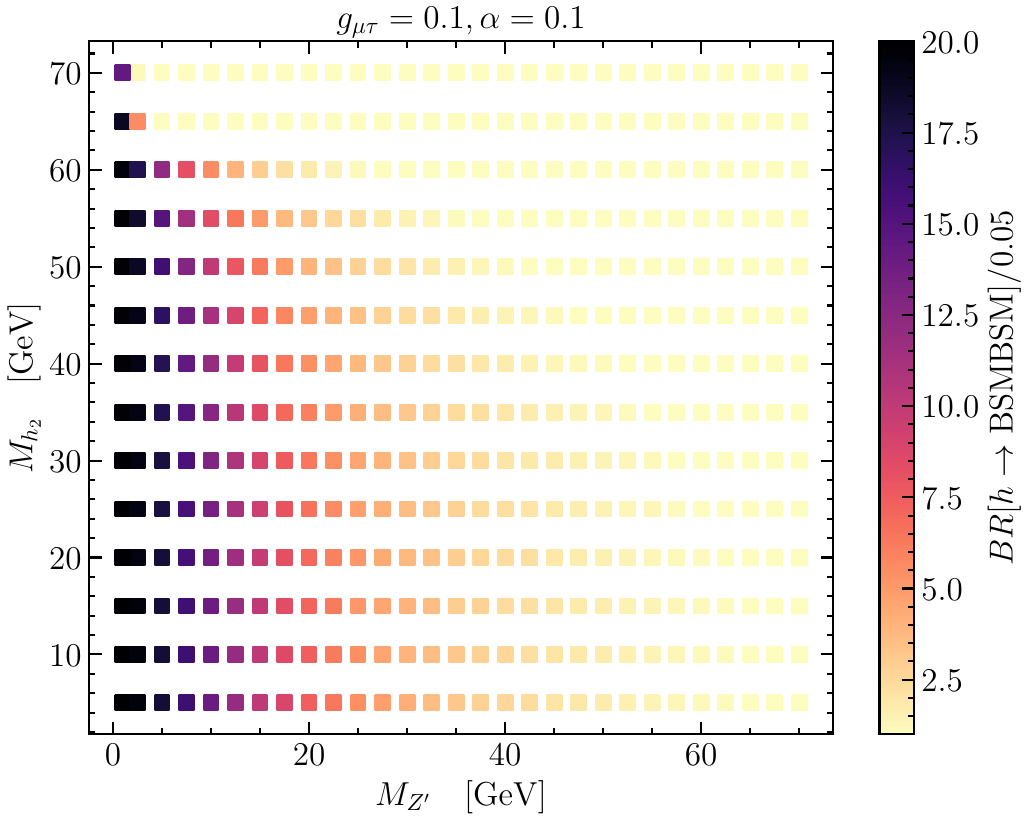}
    \includegraphics[width = 0.3\textwidth]{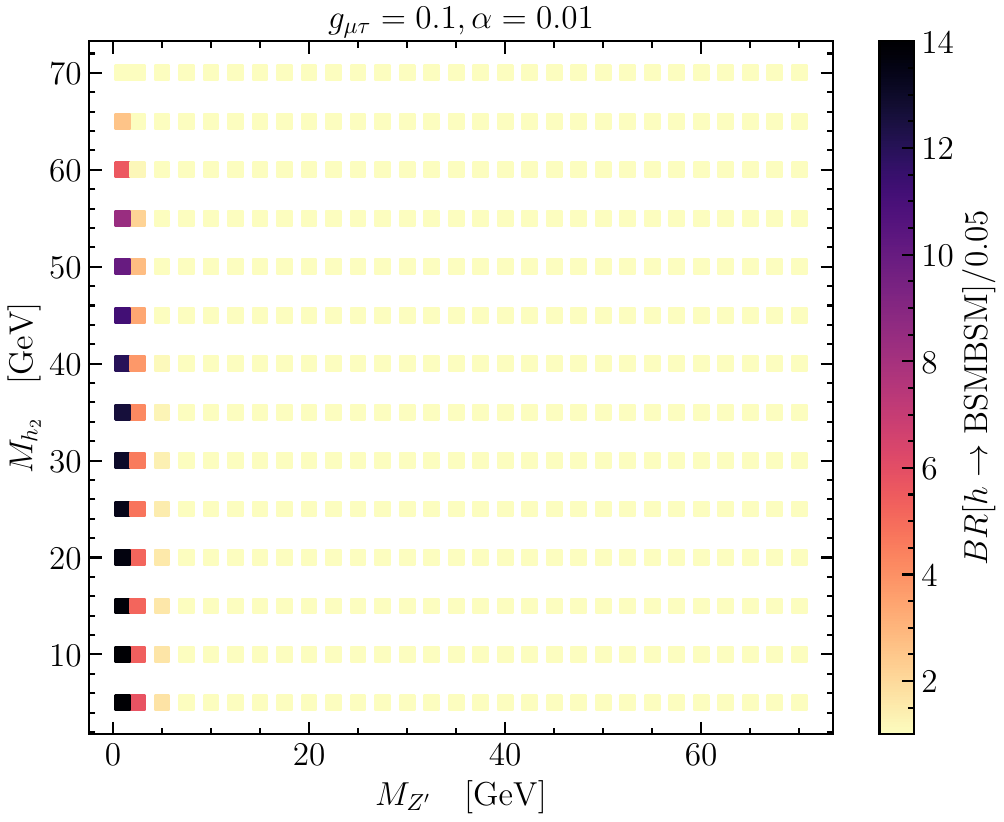}
    \includegraphics[width = 0.3\textwidth]{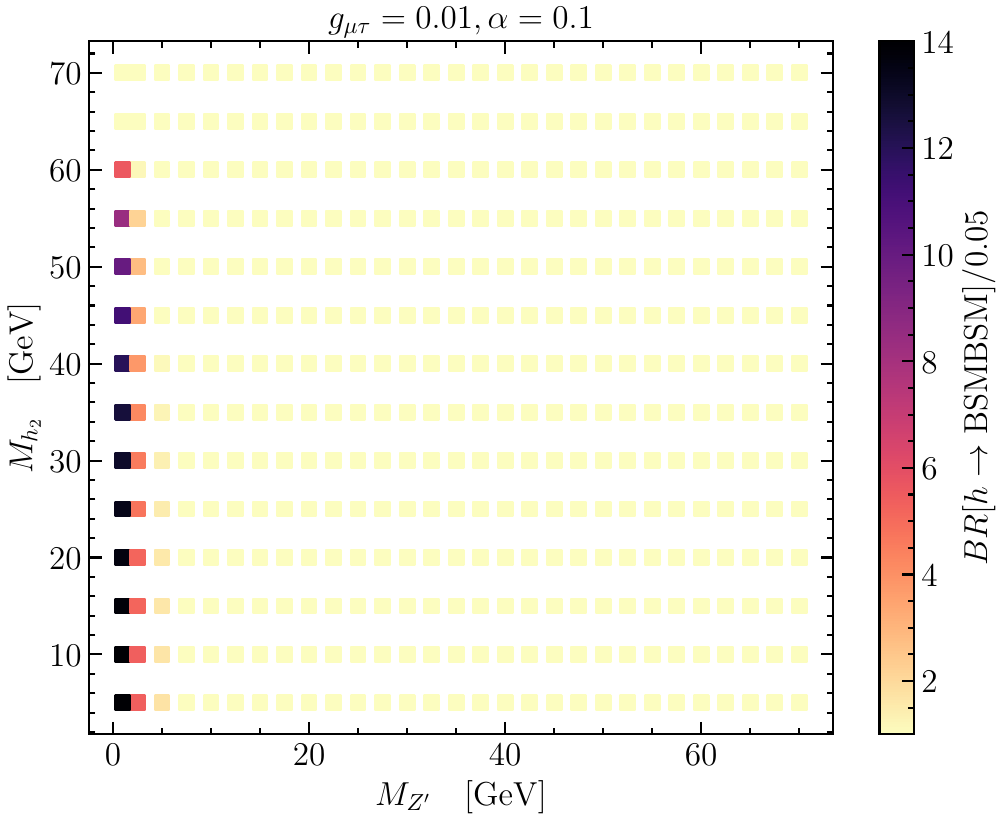}
    \caption{The branching ratio of the SM Higgs boson decay into BSM particles is colour coded and cut at 5\%, with the mass of the $Z'$ on the horizontal axis and the mass of $h_2$ on the vertical axis. The points have been computed on a grid with distance of 2.5 GeV for $M_{Z'}$ and 5 GeV for $M_{h_2}$. The combinations $g_{\mu\tau} = 0.1  \hspace{0.5em} \&  \hspace{0.5em} \alpha = 0.1$, $g_{\mu\tau} = 0.1  \hspace{0.5em} \&  \hspace{0.5em} \alpha = 0.10$, and $g_{\mu\tau} = 0.01  \hspace{0.5em} \&  \hspace{0.5em} \alpha = 0.1$ are shown from left to right.}
    \label{fig:branching_ratios}
\end{figure}

Naturally the region of interest allows for the decay of the SM Higgs boson into non-SM particles, i.e. $h_2$ and $Z'$. In order to verify that the decay of the SM Higgs boson does not a priori exclude the parameter region of interest for the 6/8 lepton search we have computed the branching ratio of $h_1\to$ BSM Particles and demand that it is not more than 5\%. The results for the $M_{Z'}-M_{h_2}$ plane can be seen in figure~\ref{fig:branching_ratios}. Notably for $g_{\mu\tau} = 0.1$ and $\alpha = 0.1$ the branching ratios are fairly significant, however lowering either $g_{\mu\tau}$ or $\alpha$ by one order of magnitude shows that the branching ratio is only relevant for low values of $M_{Z'}$. While the dependence of the branching ratio of the SM Higgs boson on $M_{Z'}$ may appear to be counter intuitive, as its decay is dominated by $h_1\to h_2 h_2$, the coupling constant for this vertex contains a $1/M_{Z'}$ term as seen in equation~\eqref{eq:feynman_rules}. When inspecting both figures~\ref{fig:exclusion_lines} and \ref{fig:branching_ratios}, it is clear that the decay of the SM Higgs boson provides no danger for the exclusion of these processes, as for low $M_{Z'}$ masses the exclusion line is well below $g_{\mu\tau} = 0.1$.

\begin{figure}[h]
    \centering
    \includegraphics[width = 0.3\textwidth]{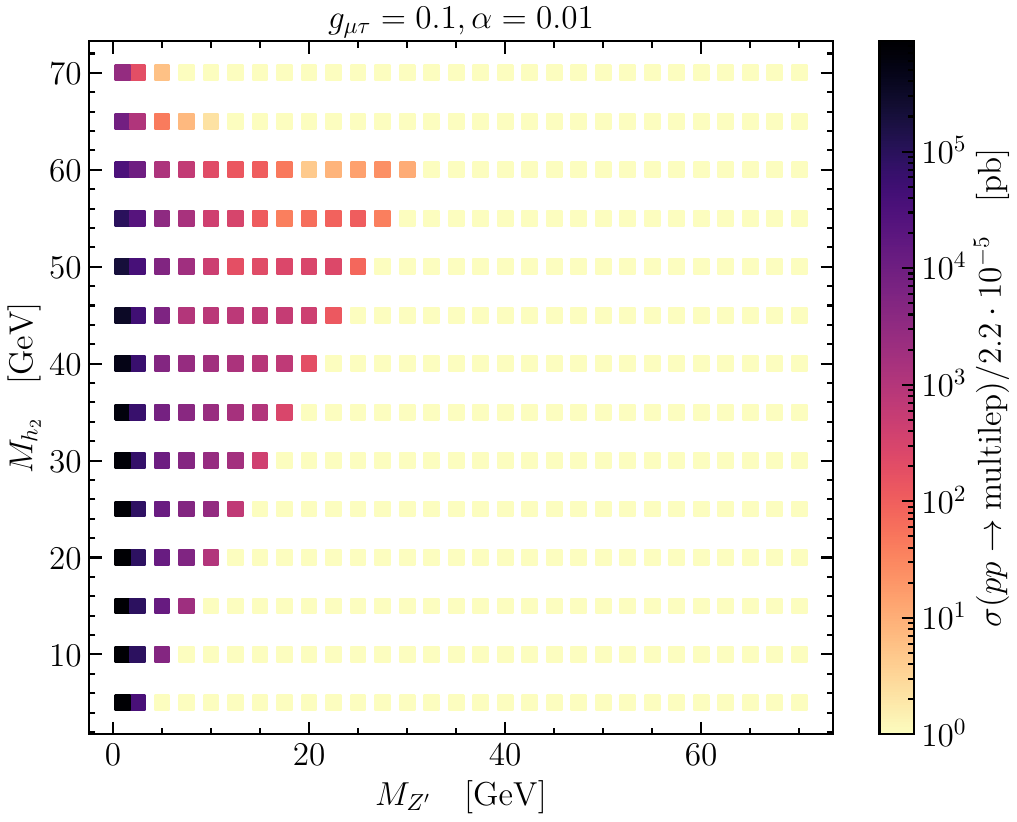}
    \includegraphics[width = 0.3\textwidth]{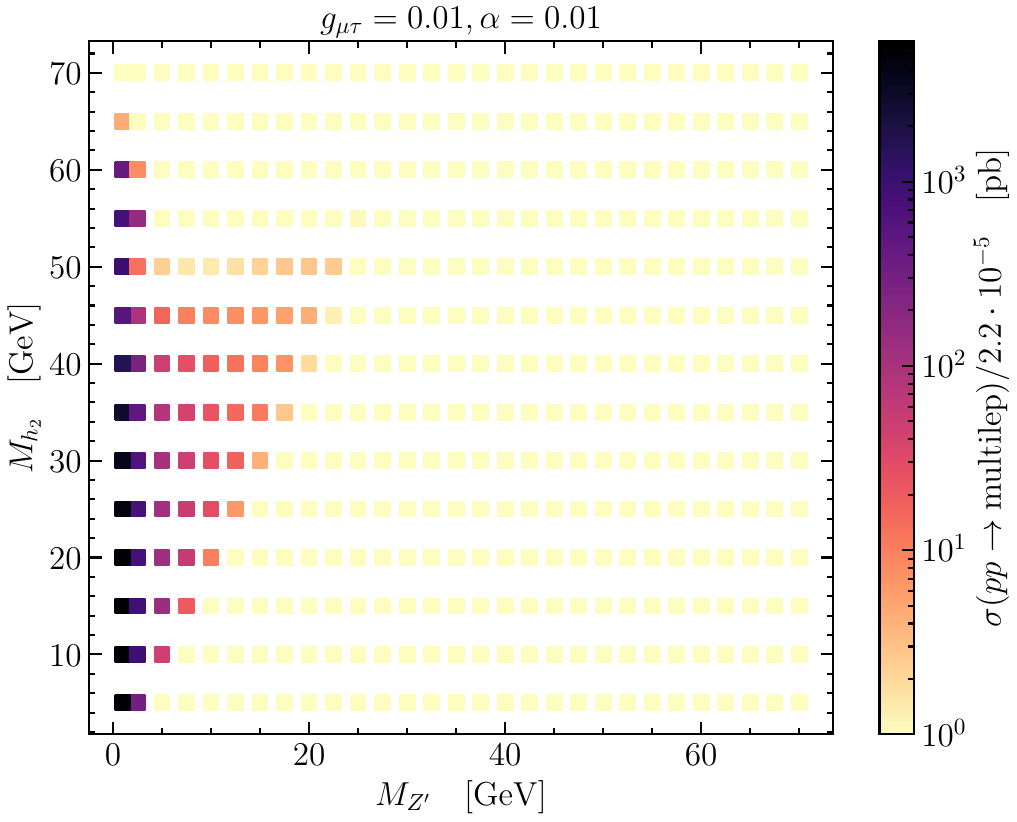}
    \includegraphics[width = 0.3\textwidth]{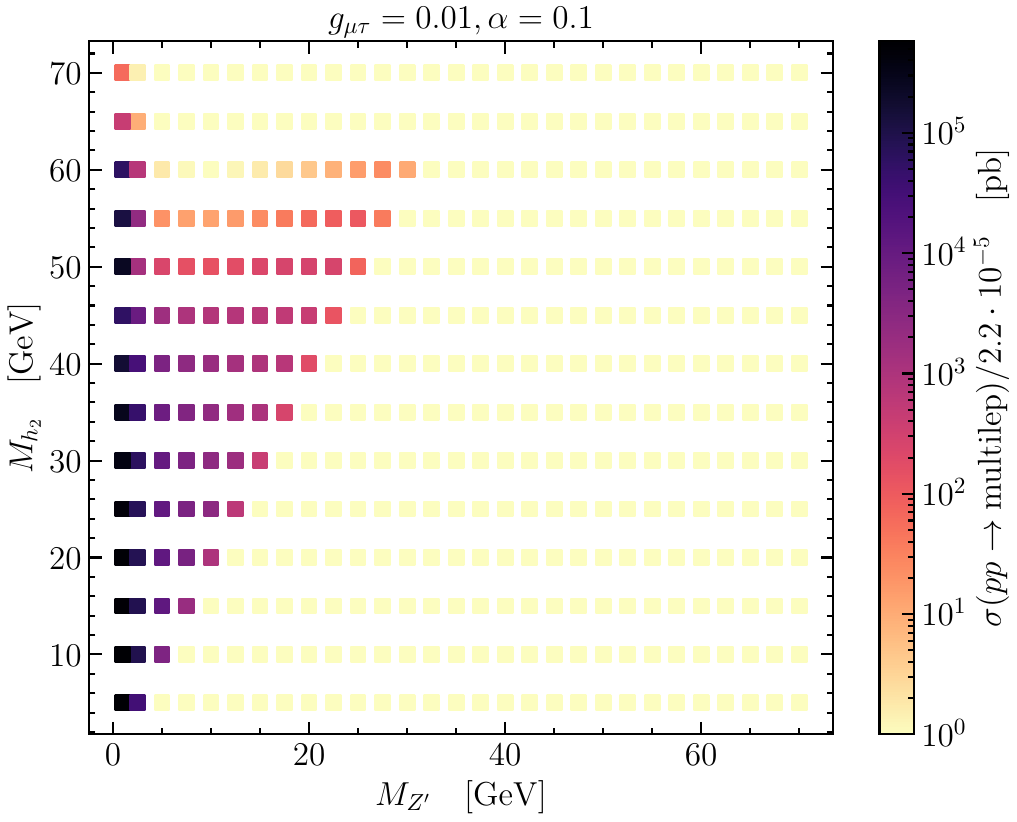}
    \caption{The cross sections of the 6/8 lepton search in colour coding cut at $2.2\cdot 10^{-5} {\rm [pb]}$ with the $Z'$ mass on the horizontal axis and the $h_2$ mass on the vertical axis. The grid has a distance of 2.5 GeV for $M_{Z'}$ and 5 GeV for $M_{h_2}$. The combinations $g_{\mu\tau} = 0.1 \hspace{0.5em}\&\hspace{0.5em} \alpha = 0.01$, $g_{\mu\tau} = 0.01 \hspace{0.5em}\&\hspace{0.5em} \alpha = 0.01$, and $g_{\mu\tau} = 0.01 \hspace{0.5em} \& \hspace{0.5em} \alpha = 0.1$ are shown on the left, middle, and right plots respectively.}
    \label{fig:cross_sections}
\end{figure}

In figure~\ref{fig:cross_sections} the cross sections of the 6/8 lepton searches are shown with a cut at $2.2\cdot 10^{-5}$ pb. Again the region of interest, namely $M_{Z'} \leq 1/2 M_{h_2} \leq 1/4 M_{h}$, is clearly visible, in addition to small values of $M_{Z'}$ for $M_{h_2} > 1/2 M_{h_1}$. For $M_{h_2} < 60$ at $g_{\mu\tau} = 0.01$ and $\alpha = 0.01$ only small values of $M_{Z'}$ are allowed. When inspecting the exclusion line corresponding to $M_{h_2} = 50$ GeV and $\alpha = 0.01$ from figure~\ref{fig:exclusion_lines} it can be seen that the line lies just below $g_{\mu\tau} = 0.01$, thus when extrapolating to $M_{h_2} = 60$ GeV the process is then indeed expected to be excluded for $g_{\mu \tau} = 0.01$. Additionally, when comparing the plots of the cross sections for $g_{\mu\tau}=0.1 \hspace{0.5em} \& \hspace{0.5em} \alpha = 0.01$ and $g_{\mu\tau}=0.01 \hspace{0.5em} \& \hspace{0.5em} \alpha = 0.1$ it can be seen that they are fairly similar for multiple values of $M_{h_2}$. This is again as expected when inspecting figure~\ref{fig:exclusion_lines}, where the exclusion line for identical values of $M_{h_2}$ but for $\alpha = 0.1$ and $\alpha =0.01$ differ approximately one order of magnitude in $g_{\mu\tau}$.

\section{Conclusion}
\label{sec:4}
In this paper we have shown that the 6/8 lepton search is a powerful probe for small values of both the $U(1)_{L_\mu - L_\tau}$ gauge coupling $g_{\mu\tau}$ and the Higgs mixing parameter $\alpha$ in the parameter region $M_{Z'} \leq 1/2 M_{h_2} \leq 1/4 M_{h_1}$, significantly outperforming the 2, 3 and 4 lepton channels. The effectiveness of this probe is compounded by the fact a small Higgs mixing parameter is able to be probed effectively in regions of the parameter space where no deviations in the SM Higgs sector are expected to be induced. However, this channel is of course limited by the size of the parameter space that is able to be probed effectively; requiring both $h_2$ and $Z'$ being able to be produced on-shell for an effective probe. Future runs of the LHC are expected to provide more stringent bounds, seeing as exclusion is directly tied to the integrated luminosity due to the clean signal provided by this channel. 
\bibliographystyle{unsrt}
\bibliography{main}

\begin{thebibliography}{10}

\bibitem{Lmu-Ltau_first_paper:PhysRevD.44.2118}
Xiao-Gang He, G.~C. Joshi, H.~Lew, and R.~R. Volkas.
\newblock Simplest ${Z}^{\ensuremath{'}}$ model.
\newblock {\em Phys. Rev. D}, 44:2118--2132, Oct 1991.

\bibitem{Scalar_DM:Baek_2016}
Seungwon Baek.
\newblock Dark matter and muon ( g - 2) in local
  {$U(1)_{L_\mu-L_\tau}$}-extended ma model.
\newblock {\em Physics Letters B}, 756:1--5, may 2016.

\bibitem{DREES2019130}
Manuel Drees, Meng Shi, and Zhongyi Zhang.
\newblock Constraints on ${U(1)_{L_\mu-L_\tau}}$ from {LHC} data.
\newblock {\em Physics Letters B}, 791:130--136, 2019.

\bibitem{Anomaly_1:PhysRev.177.2426}
Stephen~L. Adler.
\newblock Axial-vector vertex in spinor electrodynamics.
\newblock {\em Phys. Rev.}, 177:2426--2438, Jan 1969.

\bibitem{Anomaly_2:PhysRev.184.1848}
William~A. Bardeen.
\newblock Anomalous ward identities in spinor field theories.
\newblock {\em Phys. Rev.}, 184:1848--1859, Aug 1969.

\bibitem{Anomaly_cancel_1:PhysRevD.43.R22}
X.~G. He, G.~C. Joshi, H.~Lew, and R.~R. Volkas.
\newblock New-${Z}^{\ensuremath{'}}$ phenomenology.
\newblock {\em Phys. Rev. D}, 43:R22--R24, Jan 1991.

\bibitem{Anomaly_cancel_2:Ma_2002}
Ernest Ma, D.P. Roy, and Sourov Roy.
\newblock Gauged {$L_\mu-L_\tau$} with large muon anomalous magnetic moment and
  the bimaximal mixing of neutrinos.
\newblock {\em Physics Letters B}, 525(1-2):101--106, jan 2002.

\bibitem{Lmu-Ltau_coll_1:Baek_2009}
Seungwon Baek and Pyungwon Ko.
\newblock Phenomenology of {$U(1)_{L_\mu-L_\tau}$} charged dark matter at
  {PAMELA}/{FERMI} and colliders.
\newblock {\em Journal of Cosmology and Astroparticle Physics},
  2009(10):011--011, oct 2009.

\bibitem{Lmu-Ltau_coll_2:Bandyopadhyay_2009}
Priyotosh Bandyopadhyay, Sandhya Choubey, and Manimala Mitra.
\newblock Two higgs doublet type {III} seesaw with $\mu$-$\tau$ symmetry at
  {LHC}.
\newblock {\em Journal of High Energy Physics}, 2009(10):012--012, oct 2009.

\bibitem{Lmu-Ltau_LEP1:Carena_2004}
Marcela Carena, Alejandro Daleo, Bogdan~A. Dobrescu, and Tim M.~P. Tait.
\newblock {$Z^\prime$} gauge bosons at the fermilab tevatron.
\newblock {\em Physical Review D}, 70(9), nov 2004.

\bibitem{Lmu-Ltau_LEP2:Cacciapaglia_2006}
G.~Cacciapaglia, C.~Cs{\'{a}}ki, G.~Marandella, and A.~Strumia.
\newblock The minimal set of electroweak precision parameters.
\newblock {\em Physical Review D}, 74(3), aug 2006.

\bibitem{Lmu-Ltau_LHC-1:Aad_2014}
ATLAS Collaboration.
\newblock Search for high-mass dilepton resonances in pp collisions at sqrt(s)
  = 8 tev with the {ATLAS} detector.
\newblock {\em Physical Review D}, 90(5), sep 2014.

\bibitem{U(1)_Bounds:A:2023wup}
ShivaSankar~K. A., Arindam Das, Gaetano Lambiase, Takaaki Nomura, and Yuta
  Orikasa.
\newblock {Probing chiral and flavored $Z^\prime$ from cosmic bursts through
  neutrino interactions}.
\newblock 8 2023.

\bibitem{LHC_search_U1:Heeck:2011wj}
Julian Heeck and Werner Rodejohann.
\newblock {Gauged $L_\mu - L_\tau$ Symmetry at the Electroweak Scale}.
\newblock {\em Phys. Rev. D}, 84:075007, 2011.

\bibitem{LCH_search:Komachenko:1989qn}
Yu.~Ya. Komachenko and M.~Yu. Khlopov.
\newblock {On Manifestation of $Z^\prime$ Boson of Heterotic String in
  Exclusive Neutrino $N \to$ Neutrino P0 $N$ Processes}.
\newblock {\em Sov. J. Nucl. Phys.}, 51:692--695, 1990.

\bibitem{ggh_crosssec:https://doi.org/10.23731/cyrm-2017-002}
{CERN}.
\newblock Cern yellow reports: Monographs, vol 2 (2017): Handbook of lhc higgs
  cross sections: 4. deciphering the nature of the higgs sector, 2017.

\bibitem{MadGraph1:Alwall_2014}
J.~Alwall, R.~Frederix, S.~Frixione, V.~Hirschi, F.~Maltoni, O.~Mattelaer,
  H.-S. Shao, T.~Stelzer, P.~Torrielli, and M.~Zaro.
\newblock The automated computation of tree-level and next-to-leading order
  differential cross sections, and their matching to parton shower simulations.
\newblock {\em Journal of High Energy Physics}, 2014(7), jul 2014.

\bibitem{MadGraph2:Alwall_2015}
Johan Alwall, Claude Duhr, Benjamin Fuks, Olivier Mattelaer, Deniz~Gizem
  Öztürk, and Chia-Hsien Shen.
\newblock Computing decay rates for new physics theories with {FeynRules} ~and
  {MadGraph} 5{\_}{aMC}@{NLO}.
\newblock {\em Computer Physics Communications}, 197:312--323, dec 2015.

\bibitem{kleiss2021}
Ronald Kleiss.
\newblock {\em Quantum Field Theory; A Diagrammatic Approach}.
\newblock Cambridge University Press, Cambridge, 2021.

\bibitem{234_lep_search:https://doi.org/10.48550/arxiv.2109.07674}
Manuel Drees, Meng Shi, and Zhongyi Zhang.
\newblock Machine learning optimized search for the {$Z'$} from
  {$U(1)_{L_\mu-L_\tau}$} at the lhc, 2021.

\end{thebibliography}
\end{document}